\documentclass[aps,pre,showpacs,twocolumn,superscriptaddress,letterpaper]{revtex4}
\usepackage{mathrsfs}
\usepackage{amsmath}
\usepackage{bm}
\usepackage{color}
\usepackage{graphicx}
\usepackage{hyperref}

\begin{document}


\title{Global Theory to Understand Toroidal Drift Waves in Steep Gradient}
\author{Hua-sheng Xie}
\email[]{Email: huashengxie@gmail.com} \affiliation{Fusion
Simulation Center, State Key Laboratory of Nuclear Physics and
Technology, School of Physics, Peking University, Beijing 100871,
China}
\author{Bo Li}\email[Corresponding author.~]{Email: bli@pku.edu.cn}
\affiliation{Fusion Simulation Center, State Key Laboratory of
Nuclear Physics and Technology, School of Physics, Peking
University, Beijing 100871, China}

\date{\today}

\begin{abstract}
Toroidal drift waves with unconventional mode structures and
non-ground eigenstates, which differ from typical ballooning
structure mode, are found to be important recently by large scale
global gyrokinetic simulations and especially become dominant at
strong gradient edge plasmas [cf., Xie and Xiao, Phys. Plasmas, 22,
090703 (2015)]. The global stability and mode structures of drift
wave in this steep edge density and temperature gradients are
examined by both direct numerical solutions of a model
two-dimensional eigen equation and analytical theory employing
WKB-ballooning approach. Theory agrees with numerical solutions
quite well. Our results indicate that (i) non-ground eigenstates and
unconventional mode structures generally exist and can be roughly
described by two parameters `quantum number' $l$ and ballooning
angle $\vartheta_k$, (ii) local model can overestimate the growth
rate largely, say, $>50\%$, and (iii) the narrow steep equilibrium
profile leads to twisting (triangle-like) radial mode structures.
With velocity space integral, semi-local theory predicts that the
critical jump gradient of the most unstable ion
temperature gradient mode from ground state $l=0$ to non-ground
state $l=1$ is $L_T^{-1}R\sim50$. These features can have important
consequences to turbulent transport.
\end{abstract}


\maketitle

\section{Introduction and motivation}\label{sec:intro}

Drift waves\cite{Horton1999} widely exist in nonuniform magnetized
plasmas and are thought to be the dominant turbulent transport
mechanism for particle, energy and momentum. Although it has been
known theoretically\cite{Pearlstein1969,Chen1980,Horton1981} for
decades that there exists many eigenstates for toroidal drift waves,
such as ion temperature gradient mode (ITG) and trapped electron
mode (TEM), most efforts focused on the ground state branch
(`quantum number' $l=0$, aka., fundamental branch) in the past due
to that it is usually the most significant branch in the
experiments. Recently, under strong gradient edge plasma parameters,
it was found by global gyrokinetic simulations that the most
unstable branch can jump from ground state to non-ground
states\cite{Xie2015a,Xie2015b}, i.e., the most unstable branch is
non-ground states and with also usually unconventional mode
structures, whereas the conventional mode structure has typical
ballooning structure which localizes at the outboard of poloidal
plane. The modes jump and unconventional mode structures have also
been reported in local gyrokinetic
simulations\cite{Ernst2005,Wang2012}. The modes jump here is
eigenstates jump (e.g, from one TEM to another TEM) in contrast to
the jump from one kind of mode to another, such as from ITG to
TEM\cite{Rewoldt2007} and to kinetic ballooning mode
(KBM)\cite{Candy2005}. Global unconventional mode structures are
also reported in
Refs.\cite{Fulton2014,Liao2016,Xie2012,Dickinson2014}, but
are not identified as eigenstates jump.  In
Refs.\cite{Xie2015a,Xie2015b,Ernst2005,Wang2012}, the new physics
happens at steep gradient edge region ($RL_T^{-1}>40$, where $R$ is
major radius and $L_T^{-1}$ is temperature gradient scale length);
whereas most previous works study core plasma with $RL_T^{-1}<20$,
which is the reason why the unconventional mode structures and
eigenstates jump are only reported very recently.

Considering that most previous
works\cite{Hastie1979,Chen1980,Horton1981,Connor1987,Horton1999,Lu2015,Rewoldt2007,Singh2014,Abdoul2015}
of drift waves focus on the fundamental solution with weak gradient,
the non-ground solutions at strong gradient required further
theoretical studies to provide a complete picture to understand the
most general mode structure and the distributions and transitions of
eigenstates (eigenvalue solutions). The framework to understand
those recently and future gyrokinetic simulation and experimental
results should include: global solutions instead of only local
solutions, eigenstates jump, critical jump gradient (hereafter, we
discuss the critical gradient for the most unstable mode jump from
ground eigenstate to non-ground states, not the usual critical
gradient for the mode from stable to unstable), unconventional mode
structures, possible electromagnetic (EM) effects\cite{Liao2016},
consequences (e.g., to turbulent transport) and physical
understanding. These should be resolved one-by-one.

The reasons why local solutions are not adequate are mainly due to
two reasons. The first is that the mode structures from local
solution are not intuitive and may not be able to used to compared
with global simulations or experiments directly, which will also
affect the nonlinear consequences such as turbulent transport. For
example, at least a transformation (in generally, not
straightforward) to include the second dimension solution from the
1D (one dimensional) local to 2D (two dimensional) global mode
structure are required, cf. Refs.\cite{Taylor1993,Abdoul2015}. The
second reason is more important to motivate the present work: at
strong gradient the local solutions may not be
quantitatively correct and thus cannot be used to quantitatively
compare with experiments. This can be seen in a benchmark effort in
Ref.\cite{Wang2012}. We can see in Fig.14 of Ref.\cite{Wang2012}
that at weak gradient different local gyrokinetic codes can have
good agreements at real frequency and growth rate but at strong
gradient pedestal parameters the agreement breaking down. This
deviation in local codes can come from either different models
(e.g., model equation or equilibrium implementation) or the breaking
down of the local assumption. Further study is required to identify
the validation of the local model for study the strong gradient edge
parameters.

In this work, we solve a global 2D toroidal drift wave model
equation both analytically and numerically, as one step to
understand the complete picture of drift wave in steep gradient. The
results can understand several aspects of the simulations in
Ref.\cite{Xie2015a} and is an extension of the model theory in that
work, especially, which shows that non-ground eigenstates and
unconventional mode structures generally exist. Another interesting
feature in Ref.\cite{Xie2015a} (although not be emphasized there) is
twisting (triangle-like) radial mode structure. Recently, the
twisting mode have also been found in experiments and simulations
for energetic particle (EP) excited reversed shear Alfv\'en
eigenmode (RSAE)\cite{Deng2010} and beta-induced Alfv\'en eigenmode
(BAE)\cite{Zhang2010,Wang2010,Bass2013}. Global theory\cite{Ma2015}
explains that the twisting mode is due to anti-Hermitian
contributions from wave-energetic particle resonance. Without EP,
new theory is required to understand the twisting
radial mode structures in Ref.\cite{Xie2015a}. This is another
motivation of this work. We have also noticed that global 2D
numerical solutions of model drift wave equation are also reported
in Ref.\cite{Xie2012,Dickinson2014} for fundamental solution, where
modes localized at $\theta\simeq\pm\pi/2$ are found. Later we will
conclude that those solutions are merely one of series solutions and
are not the unconventional solutions in steep gradient as reported
in global simulations\cite{Xie2015a,Fulton2014}. Rotation and shear
flow\cite{Bottino2004,Dickinson2014,Abdoul2015} can also modify the
typical ballooning structure but will be neglected in the present
study.

In the following sections, Sec.\ref{sec:model} gives the model
equation and summarizes the ballooning representation theory.
Sec.\ref{sec:anly} gives the local and global analytic solutions.
Sec.\ref{sec:num2d} focuses on the global numerical solutions.
Sec.\ref{sec:gradient} studies the global gradient profile effects.
Sec.\ref{sec:kine_jump} uses a semi-local kinetic model with
velocity space integral to give a more accurate eigenstates jump
critical gradient. Sec.\ref{sec:summ} summarizes the present study.

\section{Model equation and ballooning representation}\label{sec:model}
To focus on the qualitative behavior of the general mode structures
and eigenstates, we start from a simple $i\delta$ drift wave model,
which can be used to model ion temperature gradient mode (ITG) and
trapped electron mode (TEM), and has been widely used for
theoretical studies (cf.\cite{Hastie1979,Dickinson2014,Xie2015a}).
For a large aspect ratio, circular cross section symmetric tokamak
equilibrium, the starting 2D equation (after Fourier decomposition
of toroidal direction and time dependence $\sim e^{in\zeta-i\omega
t}$) for electrostatic fluctuations potential $\delta\phi(r,\theta)$
is\cite{Hastie1979}
\begin{eqnarray}\label{eq:itg2d0}
    &\rho_s^2\frac{\partial^2\delta\phi}{\partial x^2}-b_s\delta\phi-\Big(\frac{\omega_{*e}}{\omega}\frac{\epsilon_n}{qk_\theta\rho_s}\Big)^2\Big(\frac{\partial}{\partial\theta}+
    ik_\theta s x\Big)^2\delta\phi+\\
    &\Big[\frac{\omega_{*e}-\omega(1-i\delta_e)}{\omega_{*e}\eta_s+\omega}\Big]\delta\phi-
    2\frac{\omega_{*e}}{\omega}\chi\epsilon_n\Big(\cos\theta+\sin\theta\frac{i}{k_\theta}\frac{\partial}{\partial x}\Big)\delta\phi=0,\nonumber
\end{eqnarray}
where $\rho_s\equiv \sqrt{m_iT_e}/eB$, $T_e$ is electron
temperature, $m_i$ is ion mass, $e$ is the unit charge, $B$ is the
magnetic field, $k_\theta\equiv nq/r$ is the poloidal wave number,
$n$ is toroidal mode number, $b_s\equiv k_\theta^2\rho_s^2$,
$\omega_{s}\equiv c_s/R$, $c_s\equiv\sqrt{T_e/m_i}$, $\tau\equiv
T_e/T_i$, $q\equiv rB_{\zeta}/RB_{\theta}$ is safety factor, $R$ is
major radius, $s\equiv(r/q)(dq/dr)$ is shear, $\omega_{*e}\equiv
k_\theta T_e/(eBL_n)$, $L_n^{-1}\equiv-\partial\ln n_0/dr$ is
density gradient length scale, $\epsilon_n\equiv L_n/R$,
$\eta_s\equiv(1+\eta_i)/\tau$, $\eta_i\equiv L_{T_i}^{-1}/L_n^{-1}$,
and $\delta_e$ is non-adiabtic electron response. The poloidal angle
$\theta=0$ is at the outboard mid-plane of the torus, $x=r-r_s$ is
the radial distance from the local rational surface $r=r_s$ with
$nq(r_s)=m_0$ (integer). The complex mode frequency
$\omega\equiv\omega_r+i\gamma$
($\gamma\equiv\omega_i$). In Ref.\cite{Hastie1979},
temperature gradient length scale $L_T^{-1}\equiv-\partial\ln
T_e/dr=0$, i.e., $\eta_s=1/\tau$. Parameter $\chi$ is an artificial
coupling strength, default $\chi=1$; whereas the model reduces to
cylinder drift waves at $\chi=0$. To study the equilibrium gradient
profile effect, we will let $L_n^{-1}=L_n^{-1}(x)$, which also leads
to $\epsilon_n=\epsilon_n(x)$ and
$\omega_{*e}=\omega_{*e}(x)=\omega_sk_\theta\rho_s\epsilon_n^{-1}(x)=\omega_{s0}\epsilon_n^{-1}(x)$,
where $\omega_{s0}\equiv\omega_sk_\theta\rho_s$. For simplicity, all
other parameters (say, $q=q_0$, $s$, $k_\theta\rho_s$, $\tau$,
$\eta_i$, ...) in Eq.(\ref{eq:itg2d0}) are taken independently on
radial coordinate $x$ in the following study (except in
Sec.\ref{sec:num2d}). Usually, $\delta_e$ should be an integral
operator, but we will take it as a non-negative constant here.

Using Fourier $\delta\phi(x,\theta)=\sum_mu_m(x)e^{-im\theta}$ to
rewrite Eq.(\ref{eq:itg2d0}), the 2D equation yields
\begin{eqnarray}\label{eq:itg2d}\nonumber
    &&\Big\{k^2s^2\frac{d^2}{dz^2}+\frac{1}{k^2q^2\omega^2}(z-m)^2-k^2-
    \frac{\omega(1-i\delta_e)-\epsilon_n^{-1}}{\omega+\eta_s\epsilon_n^{-1}}\Big\}u_m\\
    &&-\chi\frac{1}{\omega}\Big[(1+s\frac{d}{dz})u_{m+1}+(1-s\frac{d}{dz})u_{m-1}\Big]=0,
\end{eqnarray}
with $u_m\equiv u_m(z)=\delta\phi_m$, $z\equiv k_\theta sx=nq'x$,
$\delta m=m-m_0$, $k\equiv k_\theta\rho_s$ and $\omega$ have been
normalized by $\omega_{s0}$. The only differences from the equation
in Ref.\cite{Xie2015a} are the additional parameters $i\delta_e$,
$\tau$ and $\chi$.

Considering that usually $u_m(x)$ in Eq.(\ref{eq:itg2d}) is nearly
translational invariant under $(m,x)\to(m+1,x+\Delta r_n)$, where
$\Delta r_n=1/nq'\ll L_{eq}$ is distance between mode rational
surfaces and $L_{eq}$ is slow varying equilibrium length scale, we
can assume $u_m(x)=u_0(x-\delta m/nq')A(x)$, with
$A(x)=\bar{A}(x)e^{im\vartheta_k}$. Here, $\bar{A}(x)$ is slow
equilibrium length scale amplitude variation, and
$e^{im\vartheta_k}$ can represent the phase variation between
neighboring Fourier components. The Fourier form of radial $u_0(x)$
can be
$u_0(x)=\int_{-\infty}^{\infty}e^{im_0\eta}e^{-inq'x\eta}f(\eta)d\eta$,
thus finally we obtain the ballooning representation
\cite{Connor1978,Connor1979,Hastie1979} (assumed $n\gg1$, $q\simeq
q_0+q'x$)
\begin{eqnarray}\label{eq:bt}\nonumber
    &&\delta\phi(x,\theta)=\sum_me^{-im(\theta-\vartheta_k)}\int_{-\infty}^{\infty}d\eta
    e^{i(m-nq'x)\eta}\bar{A}(x)f(\eta)\\
    &&=A(x)\sum_me^{-im\theta}\int_{-\infty}^{\infty}d\eta
    e^{i(m-nq)\eta}f(\eta).
\end{eqnarray}
Comparing the two forms in Eq.(\ref{eq:bt}), we can see that the
$\vartheta_k$ can represent also: (a) the wave number of radial
envelope by WKB approximation $A(x)=e^{in\int\vartheta_k(x)q'dx}$
($\delta m\simeq nq'x$) and $\vartheta_k=k_r/nq'$, and (b) the
poloidal peaking angle of perturbation in 2D $(r,\theta)$ plane from
$e^{-im(\theta-\vartheta_k)}$.

Keeping only the lowest order transformation $u_m\to f(\eta)$,
$\partial/\partial z\to-i(\eta-\vartheta_k)$,
$(m-z)\to-i\partial/\partial\eta$, we can obtain a local 1D
ballooning space equation from Eq.(\ref{eq:itg2d})
\begin{eqnarray}\label{eq:itg1d_1}\nonumber
    &&\Big\{\frac{1}{q^2k^2}\frac{d^2}{d\eta^2}+\Big[\omega^2
    \frac{\omega(1-i\delta_e)-\epsilon_n^{-1}}{\omega+\eta_s\epsilon_n^{-1}}+\omega^2k^2[1+\\
    &&s^2(\eta-\vartheta_k)^2]+2\omega
    K\Big]\Big\}f(\eta,\vartheta_k)=0,
\end{eqnarray}
where $K(\eta,\vartheta_k)\equiv\chi[\cos\eta+
s(\eta-\vartheta_k)\sin\eta]$ is from the torodial
coupling terms $u_{m\pm1}$, $f(\eta,\vartheta_k)$ is the
electrostatic potential, $\eta\in(-\infty,\infty)$ is the ballooning
poloidal angle coordinate, $f(\eta)$ is Fourier transform of the
radial structure $u_0(nq-m)$. The coordinate $\eta$ can also be seen
as field line coordinate (parallel direction) due to the mapping
relation $\nabla_\parallel\delta\phi(r,\theta,\zeta,t)\to
(nq-m)u_0(x) \to
\partial_\eta f(\eta)$. Here, the ballooning angle parameter
$\vartheta_k\in[-\pi,\pi]$ [generally,
$\vartheta_k=\vartheta_{kr}+i\vartheta_{ki}$ is complex, but the
imaginary part $i\vartheta_{ki}$ can be absorbed to $\bar{A}(x)$] is
to be determined from high order\cite{Hastie1979} (2D) theory but
can be treated as a parameter in 1D model.
Eqs.(\ref{eq:itg2d0})-(\ref{eq:itg1d_1}) (or small variation) have
been studied by many
authors\cite{Connor1987,Hastie1979,Chen1980,Horton1981,Romanelli1993,Xie2012,Dickinson2014,Xie2015a}.
In this work, we will use them to study the multi-eigenstates and
global unconventional mode structures in steep gradient.

\begin{figure}
\centering
\includegraphics[width=8.0cm]{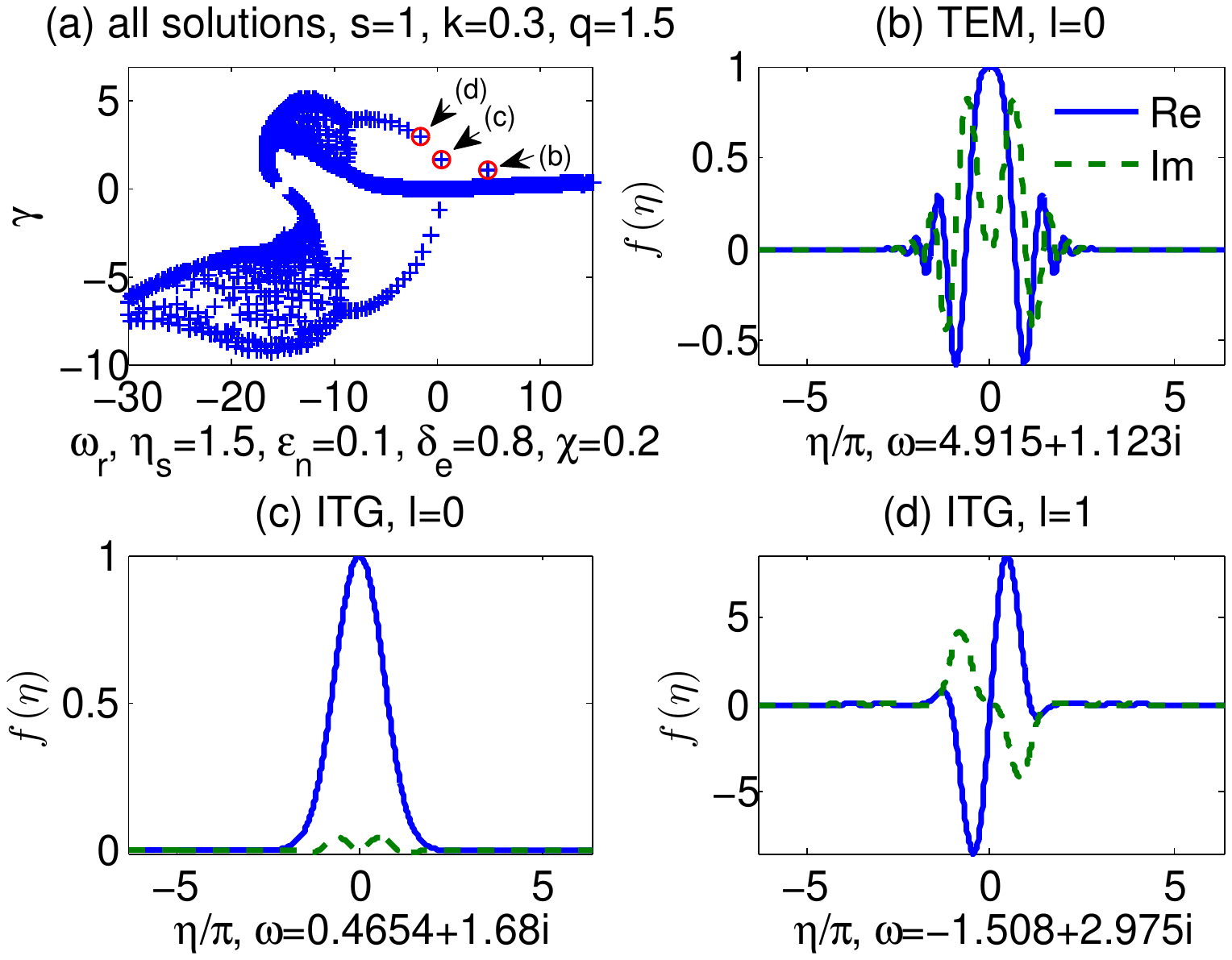}\\
\caption{The distributions of all solutions (a) of
Eq.(\ref{eq:itg1d_2}), with $s=1.0$, $k=0.3$, $q=1.5$,
$\epsilon_n=0.1$, $\eta_s=1.5$, $\delta_e=0.8$, $\chi=0.2$,
$\vartheta_k=0$ and grid parameters $N_\eta=768$,
$\eta\in[-\eta_{max},\eta_{max}]$, $\eta_{max}=20$. (b), (c), (d)
show the mode structures of TEM and ITG with quantum number $l=0,1$.
The frequencies of them are located by red circles at
(a).}\label{fig:plt_w1d_phi_Nt768}
\end{figure}

Eq.(\ref{eq:itg1d_1}) can be rewritten to polynomial form
\begin{equation}\label{eq:itg1d_2}
\{\omega^3a_3+\omega^2a_2+\omega a_1+a_0\}f=0,
\end{equation}
where $a_3=(1-i\delta_e)+k^2[1+s^2(\eta-\vartheta_k)^2]$,
$a_2=\eta_sk^2\epsilon_n^{-1}[1+s^2(\eta-\vartheta_k)^2]+2
K-\epsilon_n^{-1}$,
$a_1=\frac{1}{q^2k^2}\frac{d^2}{d\eta^2}+2\eta_s\epsilon_n^{-1} K$,
$a_0=\eta_s\frac{\epsilon_n^{-1}}{q^2k^2}\frac{d^2}{d\eta^2}$.
Eq.(\ref{eq:itg2d}) can also easily be written to polynomial form
(Appendix \ref{sec:poly_weber}). These polynomial differential
equations can be solved numerically by companion matrix
method\cite{Xie2015a} to obtain all solutions.
Fig.\ref{fig:plt_w1d_phi_Nt768} shows a typical case of the
distributions of the 1D numerical solutions of the
Eq.(\ref{eq:itg1d_2}) with central difference method and zero
boundary condition, where we take $\chi=0.2$ is to make the TEM
solution more clearly, otherwise it is usually
difficult to be distinguished from the background solutions. The
matrix system is $3N_\eta\times3N_\eta$ dimensions, thus contains
$3N_\eta$ solutions, where $N_\eta$ is grid numbers of $\eta$. We
will know later that $N_\eta$ is relevant to quantum number $l_\eta$
and $3$ means three branches, one is TEM (electron drift wave) and
another two are ITGs (but only the unstable branch is our major
interest). The positive frequency (electron diamagnetic
direction instead of ion diamagnetic direction) of ITG in
Fig.\ref{fig:plt_w1d_phi_Nt768}c is due to the modification effects
of non-zero $i\delta_e$ term. Positive ITG frequency and negative
TEM frequency have also been reported in more accurate gyrokinetic
models\cite{Dong1997,Ernst2004}. It is also readily to show that
the solutions of Eq.(\ref{eq:itg1d_2}) have several symmetric
properties: (a) For $\delta_e=0$, if $(\omega,f)$ is a solution,
$(\omega^*,f^*)$ is also a solution, where asterisk denotes complex
conjugation; (b) For $\vartheta_k=0$, the solution will be either
odd or even parity. The solutions in Ref.\cite{Xie2015a} satisfy the
above two properties. Another uncertainty is that if
$f(\eta)$ is a solution, $cf(\eta)$ is also a solution, where $c$ is
arbitrary complex number. To eliminate this
uncertainty in mode structure, say for even mode, we
can `normalize' $f\to cf(\eta)/f(0)$, with e.g., $c=1+0i$.
The normalization of odd modes in this work is slightly
arbitrary.

\section{Analytical local and global solutions}\label{sec:anly}
A physical meaningful solution should satisfy the decaying boundary
condition: for the 1D Eq.(\ref{eq:itg1d_2}) is
$f(\eta\to\pm\infty)\to0$; for the 2D Eq.(\ref{eq:itg2d}) is
$u_m(z\to\pm\infty)\to0$. Note also that for the 2D
equation, the boundary condition for $\theta$ (poloidal) direction
is periodic.

\begin{widetext}

\begin{table}[!h]
\begin{center}\caption{Analytical solutions vs. numerical solutions, parameters are same as in Fig.\ref{fig:plt_w1d_phi_Nt768}.}
\begin{tabular}{c|cccccccc}
\hline\hline $\omega$ & TEM $l=0$   &  TEM $l=1$ & ITG$^{+}$ $l=0$ &
ITG$^{+}$ $l=1$ & ITG$^{+}$ $l=2$ & ITG$^{-}$ $l=0$ & other $l=0$ &
other $l=0$ \\ \hline
Eq.(\ref{eq:itg1d_poly}) & 5.065+1.214i   &  7.647-0.586i & 0.519+2.373i & -1.143+3.413i & -2.183+3.757i & 0.601-1.694i & 4.424+5.348i & -0.770-0.020i\\
Eq.(\ref{eq:itg1d_2}) w/ apx & 5.065+1.214i   &  - & 0.519+2.373i &
-1.142+3.413i & -2.182+3.757i & 0.601-1.694i &- & -\\
Eq.(\ref{eq:itg1d_2}) w/o apx & 4.915+1.123i   &  - & 0.465+1.680i &
-1.508+2.975i & -2.737+3.393i & -0.315-1.184i  &- & - \\
\hline\hline
\end{tabular}
\end{center}\label{tab:itg1d}
\end{table}
\end{widetext}

\subsection{Analytical limit of local solution}
At analytical limit, assuming that the mode is located around
$\eta\simeq\vartheta_k$, $|\eta-\vartheta_k|\ll1$,
$K\simeq\chi[\cos\vartheta_k+(s-1)\sin\vartheta_k(\eta-\vartheta_k)+(s-1/2)\cos\vartheta_k(\eta-\vartheta_k)^2]$,
Eq.(\ref{eq:itg1d_1}) is Weber equation form
\begin{equation}\label{eq:itg1d_limit_1}
    \Big\{\frac{d^2}{d\eta^2}+g(\omega)+h(\omega)(\eta-\vartheta_1)^2\Big\}f=0,
\end{equation}
where $\vartheta_1=\vartheta_k-\frac{\chi(s-1)\sin\vartheta_k}{
k^2s^2\omega+2\chi(s-1/2)\cos\vartheta_k}$,
$g(\omega)=q^2k^2\Big\{\omega^2
    \frac{\omega(1-i\delta_e)-\epsilon_n^{-1}}{\omega+\eta_s\epsilon_n^{-1}}+\omega^2k^2+
    2\omega\chi\cos\vartheta_k-\frac{\omega[\chi(s-1)\sin\vartheta_k]^2}{
k^2s^2\omega+2\chi(s-1/2)\cos\vartheta_k}\Big\}$,
$h(\omega)=q^2k^2\big[2\omega\chi(s-1/2)\cos\vartheta_k+k^2s^2\omega^2\big]$.
The solutions are
$f=H_l(\sqrt{-h}(\eta-\vartheta_1))e^{-\sqrt{-h}(\eta-\vartheta_1)^2/2}$,
where $H_l$ is $l$-th ($l=l_\eta=0,1,2,\cdots$) Hermite
polynomial, and $g=(2l+1)\sqrt{-h}$, i.e., $g^2+(2l+1)^2h=0$, gives
\begin{eqnarray}\label{eq:itg1d_limit_3}\nonumber
    &&F(\omega)=q^2k^2\omega\Big\{\omega[\omega(1-i\delta_e)-\epsilon_n^{-1}](k^2s^2\omega+\alpha_2)
    +\\\nonumber
    &&(\omega k^2+
    \alpha_3)(\omega+\eta_s\epsilon_n^{-1})(k^2s^2\omega+\alpha_2)-(\omega+\eta_s\epsilon_n^{-1})\alpha_1\Big\}^2\\
    &&+(2l+1)^2(k^2s^2\omega+\alpha_2)^3(\omega+\eta_s\epsilon_n^{-1})^2=0,
\end{eqnarray}
where $\alpha_1=[\chi(s-1)\sin\vartheta_k]^2$,
$\alpha_2=2\chi(s-1/2)\cos\vartheta_k$,
$\alpha_3=2\chi\cos\vartheta_k$. Similar analytical solutions with
$l=0$ or $\vartheta_k=0$ are discussed in
Refs.\cite{Horton1981,Romanelli1993,Lu2015} for different purposes
of usage. Eq.(\ref{eq:itg1d_limit_3}) can be written to a seventh
order polynomial (Appendix \ref{sec:poly_weber} gives its
coefficients at $\vartheta_k=0$, which yields a fifth order
polynomial) which contains seven solutions. However, only three of
them can satisfy the decaying (depend on $\sqrt{-h}$) boundary
condition, and other explosive solutions should be dropped. The
above analytical solutions also tell us that the mode (e.g., ITGs)
can have both even and odd parities, which is determined by quantum
number $l$ (or, $l_{\eta}$) of the Hermite polynomial $H_l$. In
numerical aspect, larger $N_\eta$ is required to make larger $l$
solutions convergent. The good news is that we are mainly interested
in small $l$ solutions, which can usually be numerically handled
well. The distribution of all solutions will change for larger $l$
modes but have been convergent for small $l$ modes in
Fig.\ref{fig:plt_w1d_phi_Nt768}a, i.e., $N_\eta=768$ is sufficient
for that case.

Table I compares the analytical solutions of the approximated
equation with the numerical solutions of the original eigen
equation. We can see that with the approximation of $K$ in
Eq.(\ref{eq:itg1d_2}) the analytical solution can agree exactly with
numerical one. However, say for $l=0$, another two explosive
solutions exist in Eq.(\ref{eq:itg1d_poly}) but not in
Eq.(\ref{eq:itg1d_2}). Another difference is that the
$l=1$ TEM solution exists in Eq.(\ref{eq:itg1d_poly}) but also can
not be found (see Fig.\ref{fig:plt_w1d_phi_Nt768}) in
Eq.(\ref{eq:itg1d_2}) (however, $l=0$ TEM can be
found). This is due to the oscillation boundary condition $f\sim
e^{-i\sqrt{h}\eta^2/2}$, i.e., the numerical approach here can not
treat $\gamma\lesssim0$ TEM well. Comparing the solutions with and
without approximation of $K$ in Table I, we can find that the
analytical solutions have some deviations from the numerical ones,
which come from the deviation from the approximation that mode
localizes around $\eta\sim\vartheta_k$. The existence of an critical
gradient for the jump of most unstable mode from $l=0$ to $l\neq0$
is numerically confirmed in Ref.\cite{Xie2015a}. However,
considering of the above analytical solution, which is only rough
(with error $\gtrsim 20\%$) agreement, it seems that an analytical
expression for the critical gradient
$\epsilon_n^{c}=\epsilon_n^c(s,k,q,\cdots)$ is
challenging, let alone calculating it
in more accurate kinetic model. This is one of the reason why we
choose a simple model to do the analysis, which can yield better
general (though not accurate) understandings. Otherwise, more
powerful mathematical approaches are required. However, the major
drawback of the simplified model Eq.(\ref{eq:itg2d0}) is lacking of
Landau damping effect thus can not describe high $l$ solutions
(which usually have larger $k_\parallel$ thus will easily be damped
or stabilized) correctly. A more accurate kinetic model with
velocity space integral which contains Landau damping effect will be
discussed in Sec.\ref{sec:kine_jump}, which can give a better
prediction of the critical jump gradient.

\begin{figure}
\centering
\includegraphics[width=8.0cm]{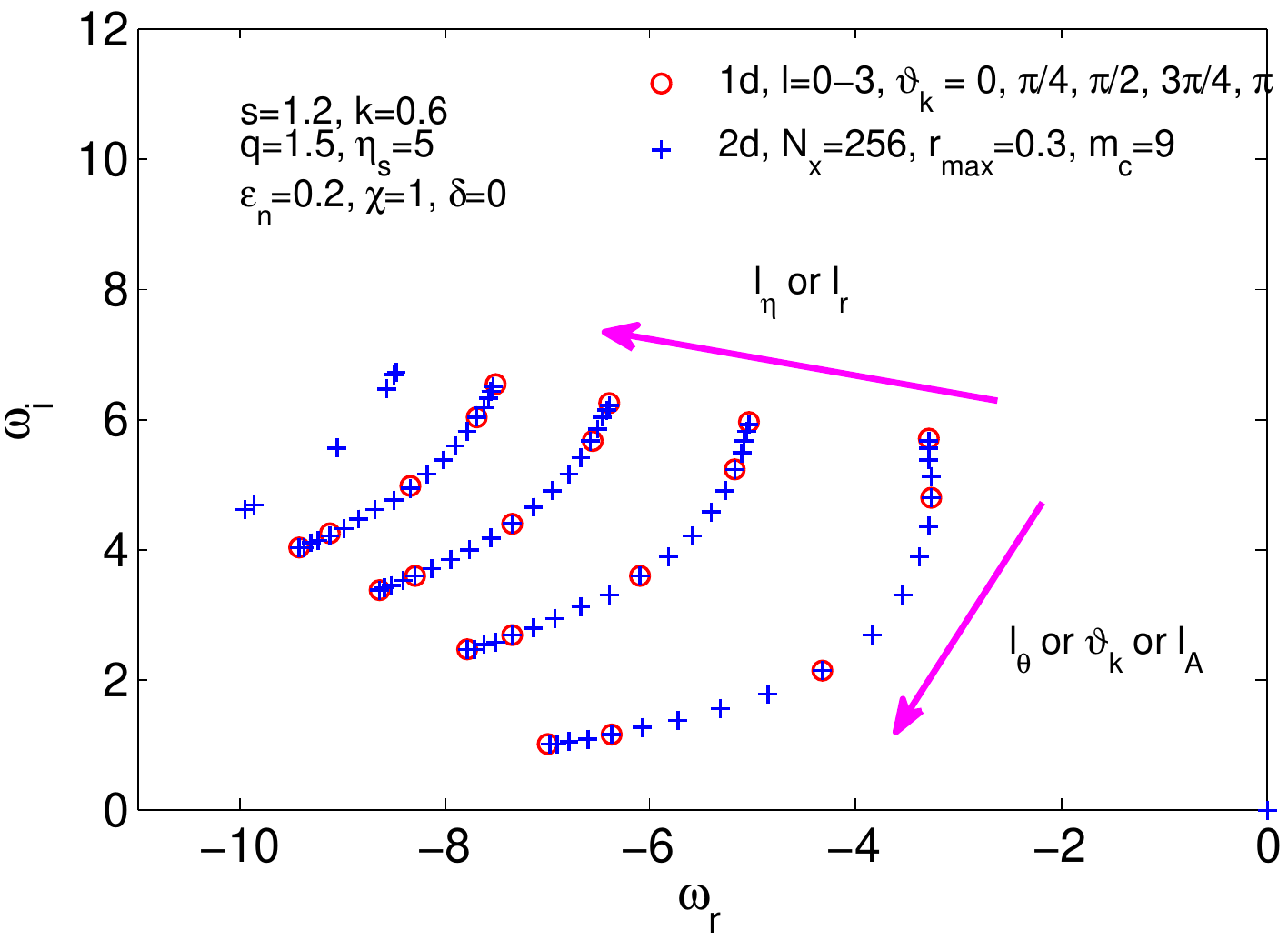}\\
\caption{Surprising relations (e.g., $l_\eta^{1d}\leftrightarrow
l_r^{2d}$ not $l_\theta^{2d}$! However, this can be explained by
below theory) between 1D and 2D solutions, agree well for
$l_\eta=0-3$, $\vartheta_k=0,\pi/4,\pi/2,3\pi/4,\pi$. The 2D
solutions are calculated directly using Eq.(\ref{eq:itg2d}), the 1D
solutions are calculated using Eq.(\ref{eq:itg1d_1}) with different
$\vartheta_k$ for each case and then gathering them together. Note
that not all solutions are shown.}\label{fig:plt_1d_vs_2d_w}
\end{figure}

\subsection{Global analytical solution}

Eq.(\ref{eq:itg1d_1}) gives local solution
$\omega=\omega(\vartheta_k,x)$. We consider the global solution with
$\partial \omega/\partial x|_{x=0}=0$. Considering higher order
ballooning representation, the second dimension
equation\cite{Hastie1979} is
\begin{equation}\label{eq:Ar}
    \frac{1}{2}\frac{\partial^2 \omega}{\partial \vartheta_k^2}\frac{d\bar{A}^2}{dx^2}+k_\theta^2s^2[\Omega-\omega(x)]\bar{A}(x)=0,
\end{equation}
where $\Omega$ is global eigen frequency, $\vartheta_k=\vartheta_m$
is the stationary position of local $\omega(\vartheta_k)$, and
usually $\vartheta_m=0,\pi$ due to symmetry. There also exists
another type of higher-order
theory\cite{Zhang1991,Xie2012,Dickinson2014,Xie2016b}, which solves
second order differential equation of $A(\vartheta_k)$ instead of
$\bar{A}(x)$. However, the final solution will be similar, thus we
will not discuss it too much. Further expanding $\omega(x)$ in
Eq.(\ref{eq:Ar}) around a stationary point, i.e., $x=0$,
$\omega(x)=\hat{\omega}+\omega_{xx}x^2/2$, yields
\begin{equation}\label{eq:Ar_1}
    \frac{1}{2}\omega_{\vartheta_k\vartheta_k}\frac{d\bar{A}^2}{dx^2}+k_\theta^2s^2[(\Omega-\hat{\omega})-\omega_{xx}x^2/2]\bar{A}(x)=0,
\end{equation}
where $\omega_{xx}\equiv\partial^2\omega/\partial x^2$ and
$\omega_{\vartheta_k\vartheta_k}\equiv\partial^2\omega/\partial
\vartheta_k^2$. This is again a Weber equation and has series
solutions $\bar{A}(x)=H_l(\sqrt{b}x)e^{-\sqrt{b}x^2/2}$,
$b=k_\theta^2s^2\omega_{xx}/\omega_{\vartheta_k\vartheta_k}$, and
\begin{equation}\label{eq:Ar_w}
    \Omega=\hat{\omega}+(l+1/2)\sqrt{\omega_{xx}\omega_{\vartheta_k\vartheta_k}}/(2k_\theta
    s),
\end{equation}
where $l=l_A=0,1,2,\cdots$ and $l_A=0$ for lowest harmonic. The
above solution will be used lately to understand the deviation of
local solution to global solution and the twisting (triangle-like)
mode structures in steep gradient. For more general cases,
Eq.(\ref{eq:Ar}) can be solved approximately by
WKB\cite{Heading1962} method.


As a first glance, we look at a special case, i.e., all profile
parameters are constants with also $\epsilon_n(x)=const.$, which
gives $\omega(x)=const.$ in Eq.(\ref{eq:Ar}). The solution can be
simple $\Omega=\omega(\vartheta_k)$ and $\bar{A}(x)$ be arbitrary.
The 1D and 2D numerical solutions of Eqs.(\ref{eq:itg1d_1}) and
(\ref{eq:itg2d}) are compared in Fig.\ref{fig:plt_1d_vs_2d_w}, where
we can find that the quantum number $l_\eta$ in 1D ballooning space
should be $l_r$ (not $l_\theta$!) in 2D real space, and the 1D
$\vartheta_k$ should be relevant to the 2D real space $l_\theta$ or
ballooning approach $l_A$. In some sense, these are surprising.
However, these can be understood from previous theory since that
$\eta$ is relevant to radial $nq(r)-m$, and $\vartheta_k$ is
relevant to $A(r)$. For the mode structures (see later), $l_\eta$
will be relevant to Fourier modes $u_m(r-r_m)$, and $l_\theta$ or
$\vartheta_k$ will be relevant to envelop $A(r)$ and poloidal
localization position $\theta$.

\begin{widetext}
\begin{center}
\begin{figure}
\centering
\includegraphics[width=17.0cm]{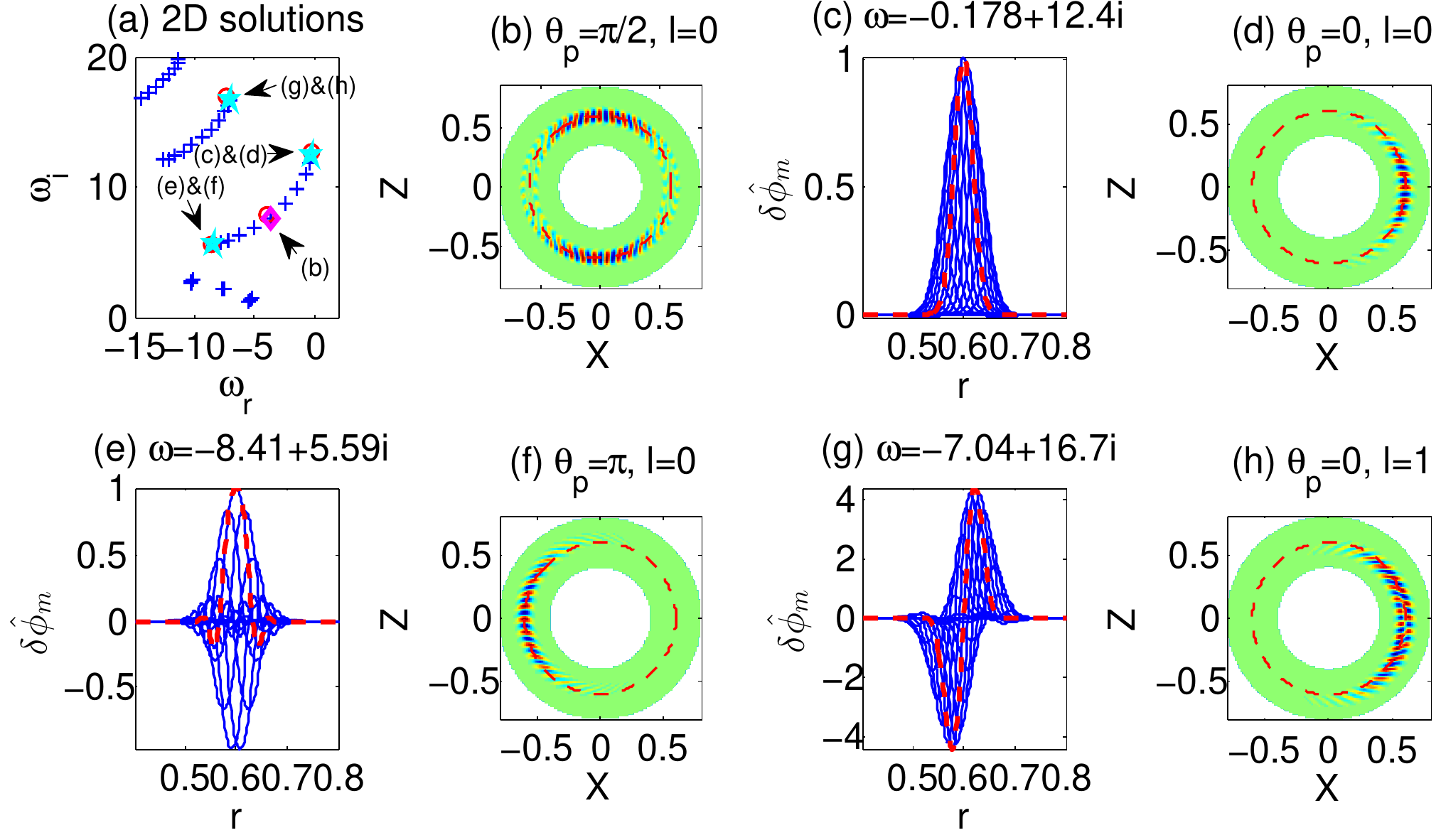}\\
\caption{Global 2D solutions, $s=2.0$, $k=0.33$, $q=1.8$,
$\epsilon_n=0.03$, $\eta_s=5.0$, $\delta_e=0.0$, $\chi=0.5$. (a)
shows series solutions, blue '+' with $m_c=5$. (b)
$\theta_p\sim\pi/2$ solution with $m_c=5$, corresponding $\omega$ is
magenta diamond in (a). (c)-(h) for real parts of $\delta\phi_m$ and
$\delta\phi(r,\theta)$ and cyan stars in (a), are three convergent
solutions with larger $m_c$. Red circles in (a) are 1D solutions.
Red dash lines in (c), (e), (g) are $\delta\phi_{m=m_0}$. Red dash
lines in (b), (d), (f), (h) are $r=r_s$, i.e.,
$x=0$.}\label{fig:plt2d_etas_x2}
\end{figure}

\end{center}
\end{widetext}

\begin{figure}
\centering
\includegraphics[width=8.0cm]{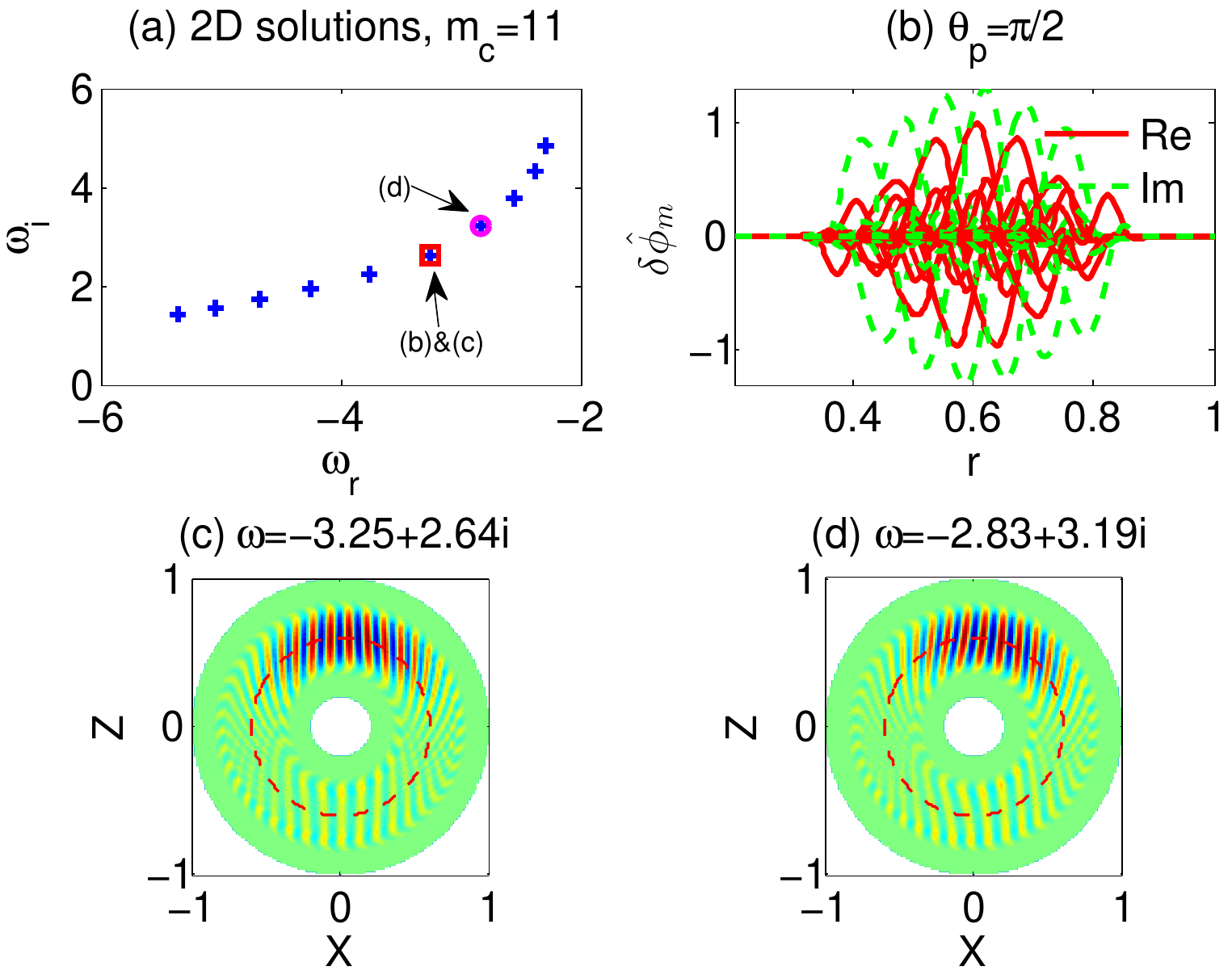}\\
\caption{Symmetry breaking from high order $O(\delta m/m)$ term
leads to single peak mode at $\theta_p=\pi/2$.}\label{fig:plt2d_xiet}
\end{figure}

\section{Global numerical solutions}\label{sec:num2d}

In the above sections, we have shown some numerical solutions to
compare with analytical solutions. In this section, we will discuss
the numerical solutions in more details. To compare with previous
works (e.g., Refs.\cite{Xie2012,Dickinson2014}), we will also adjust
Eq.(\ref{eq:itg2d}) slightly. We hope this section can provide an
overview of the global solutions, and will focus on the steep
gradient effects in next section.

\subsection{Basic features}

In numerical aspect, the discrete form of Eq.(\ref{eq:itg2dx_p3})
contains $3\times N_x\times N_m$ solutions. Again, $3$ is due to
$\omega^3$, $N_x$ is relevant to radial quantum number $l_r$, and
$N_m=2m_c+1$ is the number of poloidal $m\in[m_0-m_c,m_0+m_c]$ kept,
which is relevant to poloidal quantum number $l_\theta$. To see
these more clearly and also to compare with
Ref.\cite{Dickinson2014}, we will use $\eta_s(x)=\eta_m-\eta_gx^2$
and $\epsilon_n(x)=const.$ in this subsection.
Eq.(\ref{eq:itg2dx_p3}) can be used directly except change $\eta_s$
to $\eta_s(x)$.

Fig.\ref{fig:plt2d_etas_x2} shows solutions of a typical global 2D
case. Local parameters $s=2.0$, $k=0.33$, $q=1.8$,
$\epsilon_n=0.03$, $\eta_s=5.0$, $\delta_e=0.0$, $\chi=0.5$; global
parameters $n=20$, $r_s=0.6$, $\eta_g=1.25n$; grid parameters
$N_x=256$. Most solutions in panel (a) and the mode structure in
panel (b) will change if we use larger $m_c$, e.g., $m_c=5\to13$,
which means that those solutions are not convergent. This is not
surprising because the 2D $m_c$ determines the number of solutions
between the 1D $\vartheta_k=0$ to $\vartheta_k=\pi$. The
$\theta_p\sim\pm\pi/2$ solution has two peaks at both
$\theta\sim\pi/2$ and $-\pi/2$, whereas 1D theory of
$\vartheta_k=\pi/2$ will give only one peak. Three convergent
solutions are also shown, which agree well with ballooning 1D and 2D
theory in previous sections for both frequency and mode structure,
e.g., $\vartheta_k=0,\pi$ solutions localize at $\theta_p\sim0,\pi$
and $l=1$ solution has odd parity of $\delta\phi_m(x)$ and
$\vartheta_k=\pi$ solution has different envelop or phase for
different $\delta\phi_m$.

Considering that panel (b) solution is not convergent, it is not
known yet whether global solutions can really contains arbitrary
$\vartheta_k$ solutions as in Fig.\ref{fig:plt_1d_vs_2d_w}. However,
this solution still tell us that the mode structure is symmetric for
$\theta=0$. This feature is similar as observed in some solutions in
Ref.\cite{Xie2015a}, i.e, usually mode structure will peak at both
$\theta_p\sim\vartheta_k$ and $-\vartheta_k$.

Up to this step, we can give a short summary of basic features of
unconventional modes: $\vartheta_k$ leads to quasi-continuous change
of frequency and envelop or phase of mode structure, $l_\eta$ leads
to discontinuous jump of frequency and high order harmonic of
$\delta\phi_m$. The observations of global solutions in
Refs.\cite{Fulton2014,Xie2015a} can be combination of both effects
of $\vartheta_k$ and $l_\eta$, depending on the mode
structure feature and how the frequency changes (quasi-continuous or
discontinuous). Why (especially the physics behind) the most
unstable solutions will change from $\vartheta_k=0$ and $l_\eta=0$
to $\vartheta_k\neq0$ and $l_\eta\neq0$ is out of the scope of this
work, and required further study.

\subsection{Symmetry breaking}
There exists many different sources for poloidal symmetry breaking
(or, up-down asymmetry) of the mode structure, i.e., away from
$\theta=0$, such as rotation flow and equilibrium profile. For
example, Refs.\cite{Dickinson2014, Abdoul2015} discuss the solutions
with poloidal peaking $\theta\neq0$ with linear profile
$\eta_s=\eta_m-\eta_gx$ and toroidal shear flow, where poloidal
tilting mode structures are shown. The tilting structure can be
understood from ballooning theory as that $\vartheta_k=\vartheta_m$
where $\vartheta_m$ is slightly away from $0$.

In this subsection, as an example, we study the symmetry breaking
from a higher order term of $(1/r^2)(\partial^2/\partial\theta^2)$,
i.e., treating it as $(m/r)^2(1+\delta m/m_0)^2$ as in
Refs.\cite{Zhang1991,Xie2012} instead of $(m/r)^2$ in
Sec.\ref{sec:model}. The main purpose of this subsection is to show
the single peak solution of $\vartheta_k\neq0$.

By keeping an additional $O(\delta
m/m)$ term and dropping $i\delta_e$ term, the 2D
Eq.(\ref{eq:itg2d}) changes to\cite{Xie2012}
\begin{eqnarray}\label{eq:itg2d_o}\nonumber
    &&\Big\{k^2s^2\frac{d^2}{dz^2}+\frac{1}{k^2q^2\omega^2}(z-m)^2-k^2-
    \frac{\omega-\epsilon_n^{-1}}{\omega+\eta_s\epsilon_n^{-1}}\\\nonumber
    &&+\underbrace{[\frac{k^2}{\omega+\eta_s\epsilon_n^{-1}}+\frac{\omega-\epsilon_n^{-1}}{(\omega+\eta_s\epsilon_n^{-1})^2}]\epsilon_n^{-1}\eta_s\frac{\delta m}{m}}_{O(\delta m/m)}\Big\}u_m\\
    &&-\chi\frac{1}{\omega}\Big[(1+s\frac{d}{dz})u_{m+1}+(1-s\frac{d}{dz})u_{m-1}\Big]=0.
\end{eqnarray}
The above equation can yield a similar equation as
Eq.(\ref{eq:itg2dx_p3}) and can be numerical solved in a similar
manner, but with a more $\omega^4$ term.

Direct numerical solutions of the global
Eq.(\ref{eq:itg2d_o}) shown in Fig.\ref{fig:plt2d_xiet}
confirm the ballooning theory of the second kind
solution in Ref.\cite{Xie2012}, with parameters $s=1.2$, $k=0.6$,
$q=1.5$, $\epsilon_n=0.1$ and $\eta_s=3.0$. However, our solution
$\theta_p=\pi/2$ whereas $\theta_p=-\pi/2$ in Fig.5(a) of
Ref.\cite{Xie2012}, which is due to a sign difference in the
equation (i.e., set $\chi=-1$ in our equation can reproduce the
equation in Ref.\cite{Xie2012}). We can also see in
Fig.\ref{fig:plt2d_xiet}(d) that there also exist series other
solution with peaking just slightly away from $\theta_p=\pi/2$.
However, in Fig.\ref{fig:plt2d_xiet}, only the red square
$\theta_p=\pi/2$ solution is convergent and other solutions will
change if we use larger $m_c$. It is not clear whether
those unconvergent solutions can physically exist. The reason why
also asymmetry mode exist in Ref.\cite{Xie2015a} [e.g.,
Fig.2(d),(e),(h),(i)] under strong gradient without apparent
symmetry breaking source (e.g., rotation, flow) still need further
study. Refs.\cite{Migliano2013,Singh2014} have studied the finite
ballooning angle $\vartheta_k$ effects via gyrokinetic model which
are possible to provide some hints to understand the source of
symmetry breaking.

To this step, the unconventional mode structure with single peak
$\theta\simeq\pi/2$ or $-\pi/2$ in Refs.\cite{Dickinson2014,Xie2012}
are not the one reported in Ref.\cite{Xie2015a}. Because that the
solutions in Refs.\cite{Dickinson2014,Xie2012} is still ground state
$l=0$ and the unconventional structures come only from
$\vartheta_k\neq0$. However, the simulation results in
Ref.\cite{Xie2015a} usually show multi-peak and eigenstates jump
($l\neq0$), and do not have the apparent symmetric breaking sources
[e.g., linear $\eta_s(x)=\eta_m-\eta_gx$ profile or the same high
order term as in Eq.(\ref{eq:itg2d_o})] as in
Refs.\cite{Dickinson2014,Xie2012}.

\begin{figure}
\centering
\includegraphics[width=6.0cm]{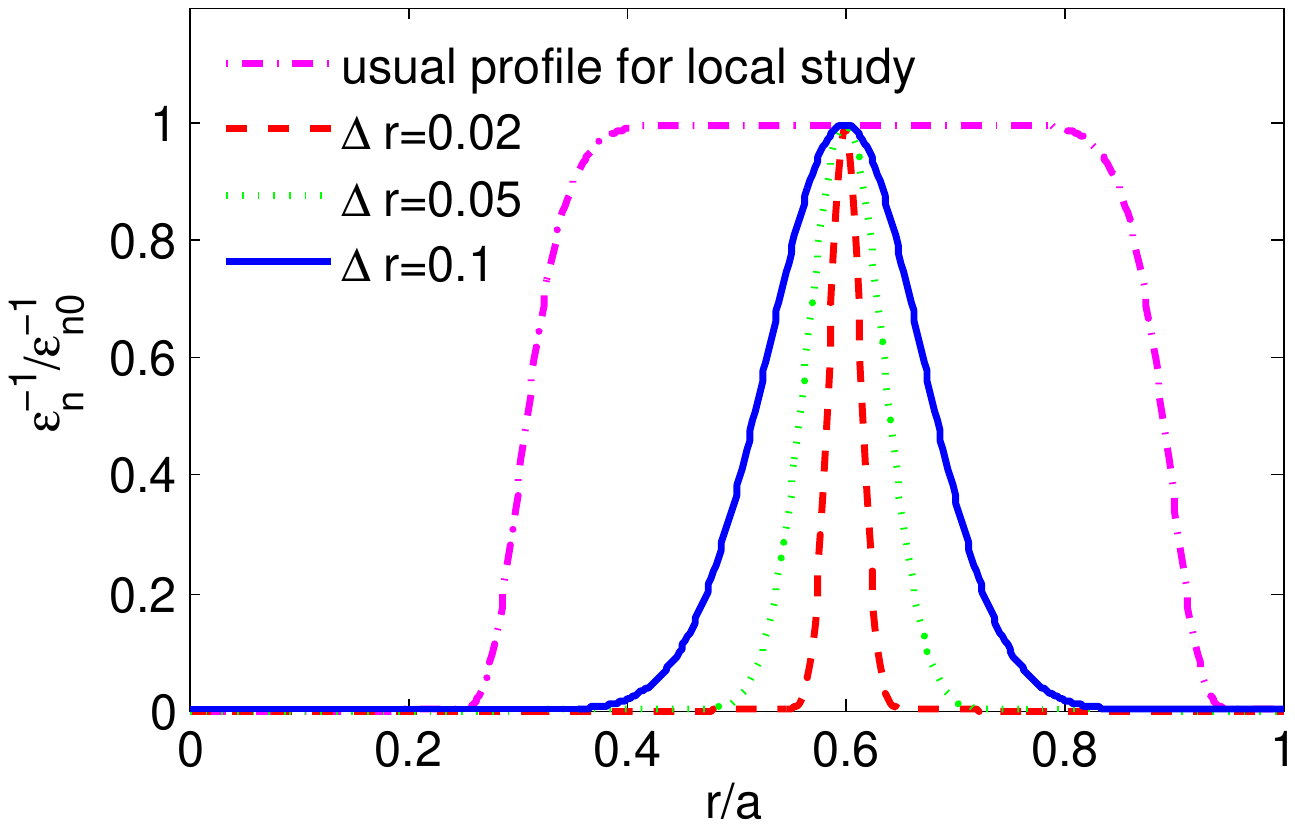}\\
\caption{The $\epsilon_n^{-1}=\epsilon_{n0}^{-1}e^{-(r-r_s)^2/\Delta
r^2}$ used for study of the global gradient profile effects. The
magenta dash-dot line shows usual used profile in global code to
mimic local model.}\label{fig:plt_epsn_r}
\end{figure}

\begin{figure}
\centering
\includegraphics[width=9.0cm]{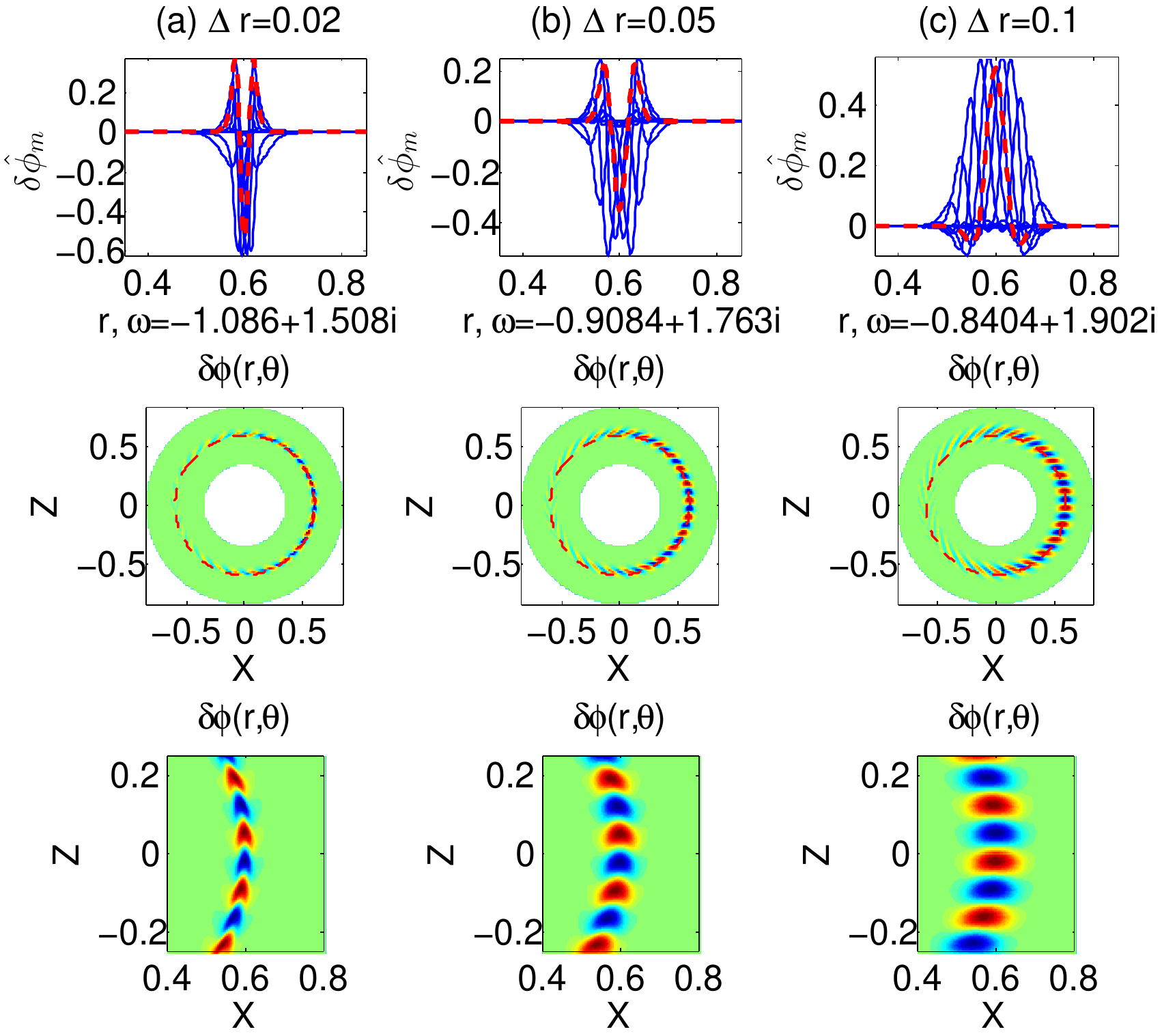}\\
\caption{Gradient profile effect
$\epsilon_n^{-1}=\epsilon_{n0}^{-1}e^{-(r-r_s)^2/\Delta r^2}$, scan
$\Delta r$. Twisting (triangle-like) mode structure for smaller
$\Delta r$.}\label{fig:scan_Dr_mode}
\end{figure}

\begin{figure}
\centering
\includegraphics[width=8.0cm]{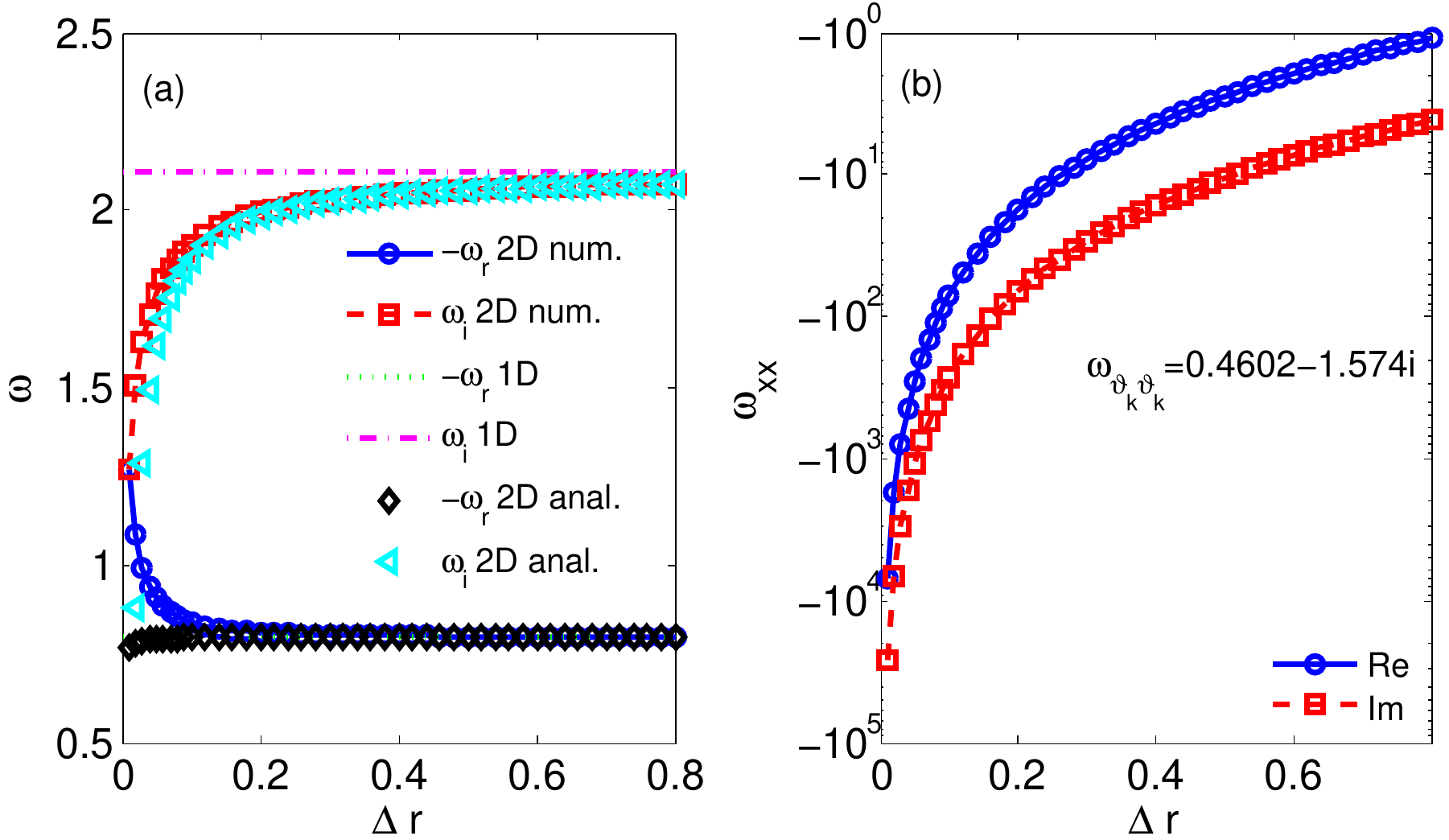}\\
\caption{Gradient profile effect
$\epsilon_n^{-1}=\epsilon_{n0}^{-1}e^{-(r-r_s)^2/\Delta r^2}$, scan
$\Delta r$. 1D v.s. 2D. Smaller $\Delta r$, smaller growth
rate.}\label{fig:scan_Dr_wr_add}
\end{figure}

\begin{figure}
\centering
\includegraphics[width=8.0cm]{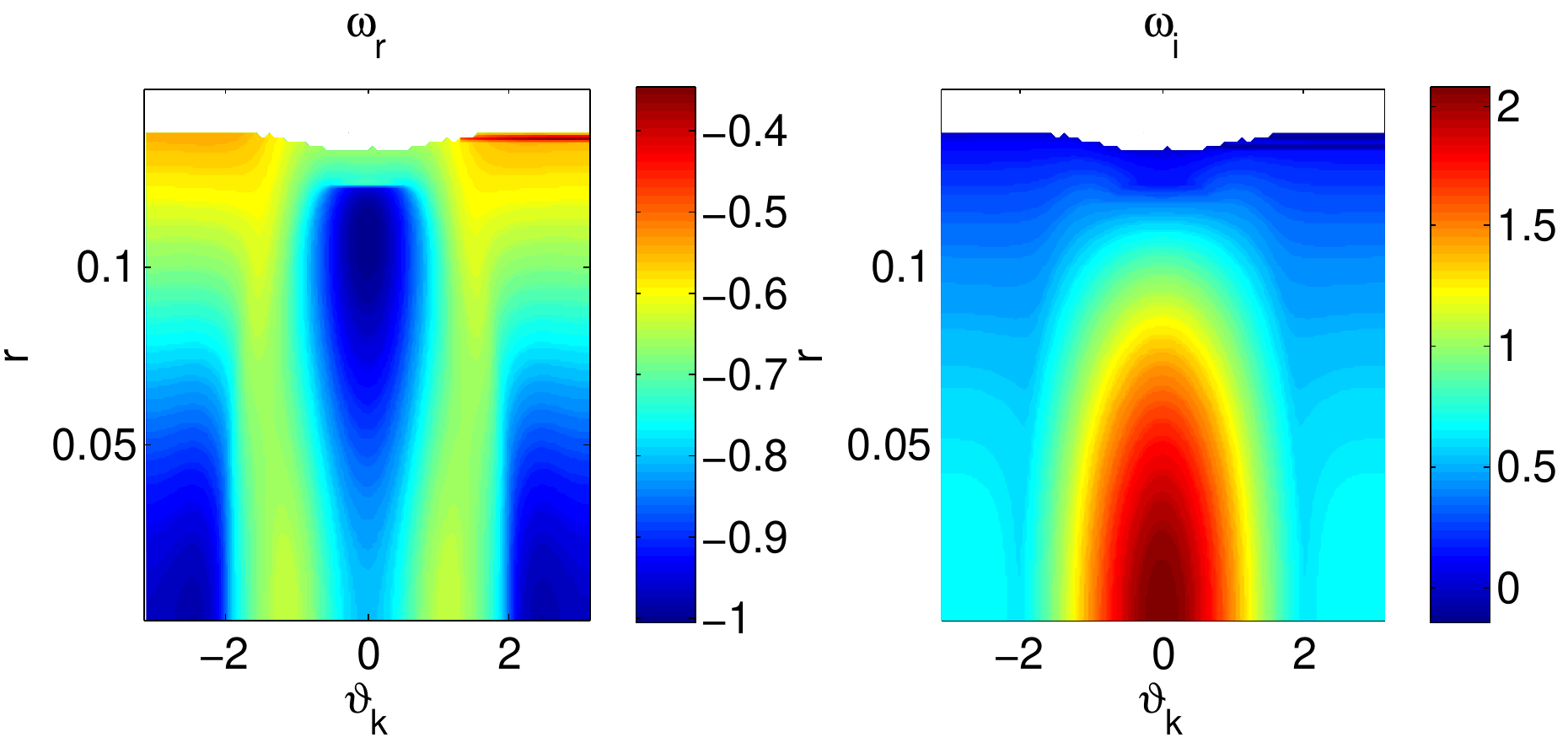}\\
\caption{Scan 1D $\omega(x,\vartheta_k)$, where the maximum growth
rate is at $(x=0,\vartheta_k=0)$.}\label{fig:plot_1D_w_xt}
\end{figure}

\begin{figure}
\centering
\includegraphics[width=8.0cm]{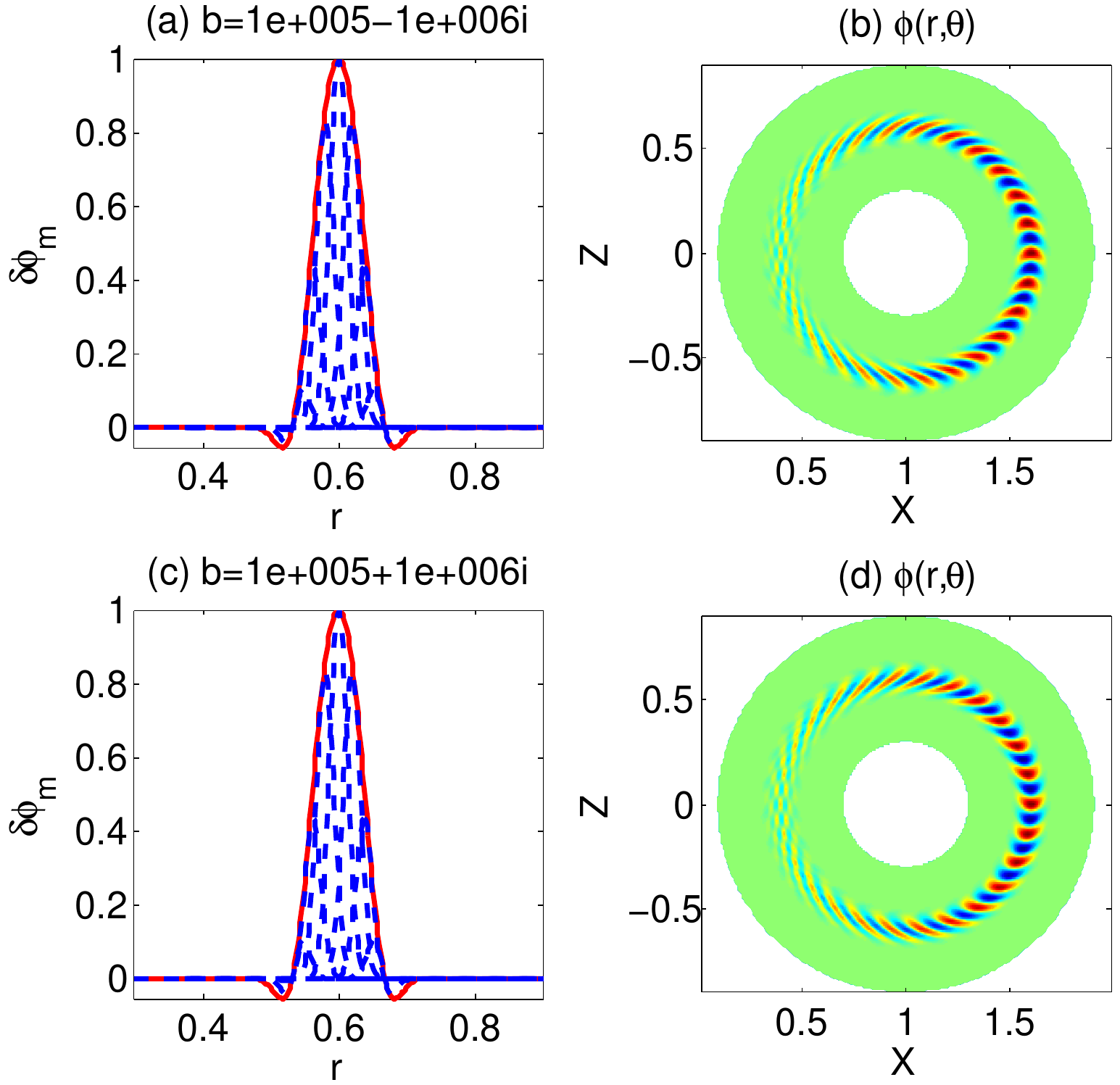}\\
\caption{The imaginary part of the parameter
$b=k_\theta^2s^2\omega_{xx}\omega_{\vartheta_k\vartheta_k}$
determines the twisting direction (clockwise or
anti-clockwise).}\label{fig:phim2phirt_1}
\end{figure}

\section{Gradient profile effect}\label{sec:gradient}
We study the equilibrium effect from gradient profile
$L_n^{-1}=L_n^{-1}(x)$, which is more analogous to the cases of
global gyrokinetic simulations in Ref.\cite{Xie2015a} by GTC code.
To model the global steep profile, we take
$\epsilon_n^{-1}=\epsilon_{n0}^{-1}e^{-(r-r_s)^2/\Delta r^2}$
(Fig.\ref{fig:plt_epsn_r}), where $\Delta r$ determines the width of
the steep profile region. At strong gradient tokamak edge plasmas,
$\Delta r$ can small to $0.01-0.05$.

The framework of analytical solutions has been discussed in
Sec.\ref{sec:anly}. In this section, we will focus on the numerical
solutions and compare them with analytical theory.
Fig.\ref{fig:scan_Dr_mode} shows typical direct numerical solutions
(only fundamental $l=0$ solutions are shown) of
Eq.(\ref{eq:itg2dx_p3}) with same local parameters ($s=1.0$,
$k=0.3$, $q=2.6$, $\epsilon_n=0.8$, $\eta_s=2.0$, $\delta_e=0.0$,
$\chi=1.0$, $n=10$, $r_s=0.6$) but different $\Delta r$, i.e.,
$0.02$, $0.05$ and $0.1$. We can see that smaller $\Delta r$ gives
smaller growth rate $\gamma$ and also more twisting (triangle-like)
of the mode structure.

These deviations of frequency and twisting of mode structure can be
well understood from previous analytical theory. From local and
global theory, smaller $\Delta r$ gives larger $\Omega_{xx}$ and
thus larger $\Omega-\hat{\omega}$, which explains the deviation of
frequency. The solutions are shown in Fig.\ref{fig:scan_Dr_wr_add}
and also are compared with numerical global 2D solutions. We see
that analytical $\gamma$ can agree well with numerical one at
$\Delta r>0.15$. For small $\Delta r$, the approximation used in the
analytical theory will not be sufficient.
Fig.\ref{fig:scan_Dr_wr_add} shows that the local model can differ
$50\%$ with global model at $\Delta r=0.01$.
Fig.\ref{fig:plot_1D_w_xt} shows the 1D local solutions
$\omega(x,\vartheta_k)$, which justifies the assumption in
calculating the global 2D solutions in Sec.\ref{sec:anly}, i.e., the
maximum growth rate is stationary at
$(x=0,\vartheta_k=0)$. The twisting mode structure is identified to
come from the imaginary part of parameter
$b=k_\theta^2s^2\omega_{xx}\omega_{\vartheta_k\vartheta_k}$ [e.g.,
for $\Delta r=0.04$, we have $\omega_{xx}=(-0.44-1.66i)\times10^3$,
which yields $b=(1.69-1.02i)\times10^6$]. To show the influence of
$b$ to the twisting mode structure more clearly, we use
$b=(0.1\pm1.0i)\times10^6$ to plot the global mode structures, which
are shown in Fig.\ref{fig:phim2phirt_1} [panels (a) and (c) are
difficult to distinguish, but the differences between (b) and (d)
are clear]. We see that in panels (a) and (b), the twisting
direction is anti-clockwise, which agrees with
Fig.\ref{fig:scan_Dr_mode}; whereas in panels (c) and (d) the sign
of the imaginary part of $b$ is changed from $'-'$ to $'+'$, which
leads to the twisting direction to clockwise. If $Im(b)=0$, twisting
structures vanish. Thus, the theory can also explain why ideal
ballooning mode (IBM) in steep gradient does not have twisting
radial structure, because that $\omega^{\rm IBM}=i\gamma$ and
$Re(\omega^{\rm IBM})=0$. The reason why ideal Alfv\'en eigen modes
without EP driven do not have twisting mode structure is similar,
i.e., the eigen frequency $\omega$ has only real part. Considering
that $k_\theta^2s^2$ and $\omega_{\vartheta_k\vartheta_k}$ are
determined by local parameters, the global profile mainly affects
$\omega_{xx}$: smaller $\Delta r$ leads to larger $\omega_{xx}$ and
then larger $Im(b)$. Thus, the twisting mode structures is a global
effect which is not included in local model.

In summary, the steep gradient can lead to at least two significant
global effects: deviation of the frequency and twisting of the mode
structure. These global effects can not be handled well by local
models and thus one should be careful in using local model to
understand physics or to explain experimental observations.

\section{Eigenstates jump with velocity space
integral}\label{sec:kine_jump} The present work is to provide more
analytical insights to understand the eigenstates jump and the
unconventional mode structures of drift waves. In the previous
sections, we have investigated the model equation in more details
than those in Ref.\cite{Xie2015a}. However, a very important issue,
i.e., the critical gradient $\epsilon^c$, is still not examined
carefully. The critical gradient is confirmed to exist in
Ref.\cite{Xie2015a} by the model equation. Here, we consider the
case with Landau damping and with more accurate finite Larmor radius
(FLR) effects, via including of the velocity space
$(v_\parallel,v_\perp)$ integral.

\begin{figure}
\centering
\includegraphics[width=9.0cm]{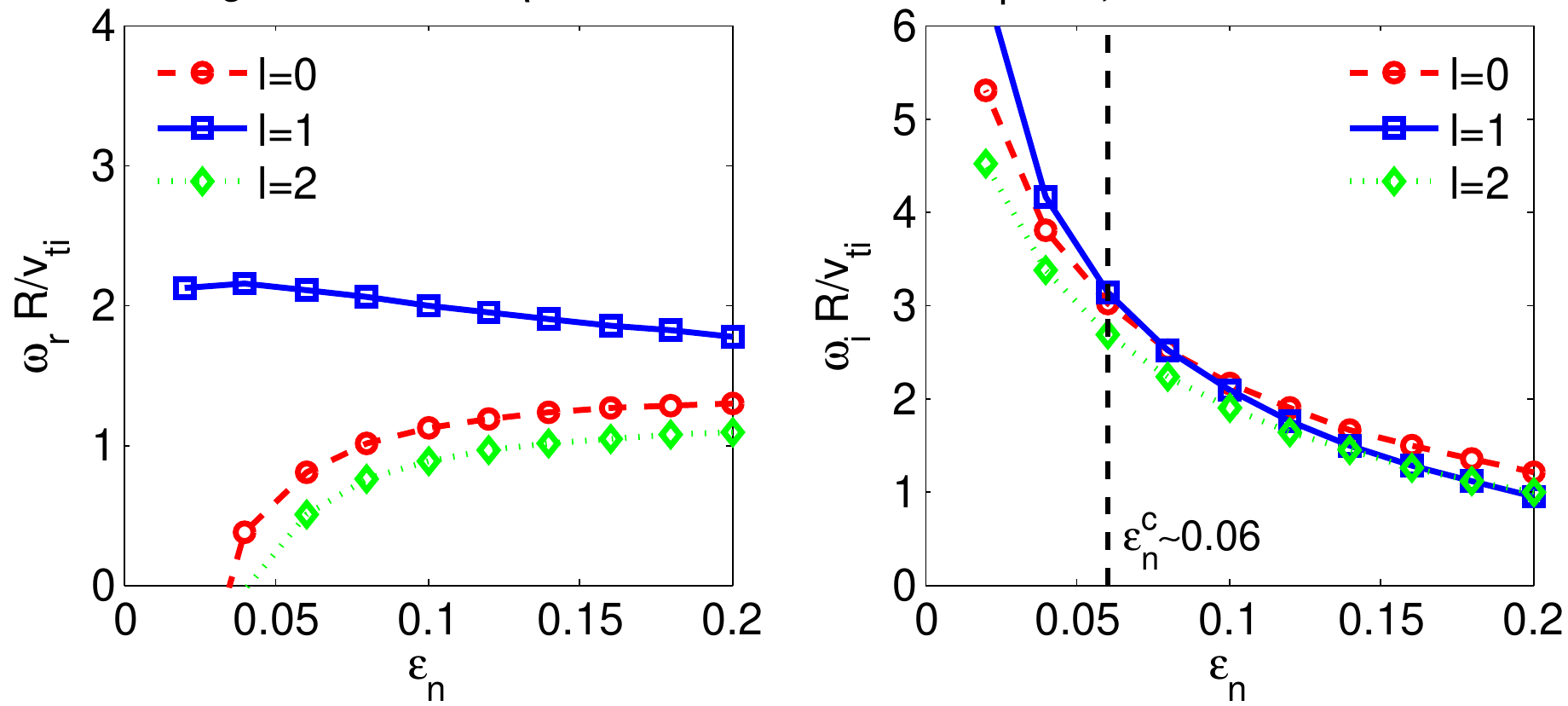}\\
\caption{Semi-local kinetic ITG dispersion relation shows the
critical gradient jump of the most unstable mode from $l=0$ to $l=1$
when $\epsilon_n<\epsilon^c\sim 0.06$. Parameter
$k_\theta\rho_i=0.4$, $\eta_i=3.114$, $\tau=1$, $s=0.78$ and
$q=1.4$.}\label{fig:kinetic_jump}
\end{figure}

We only consider ITG with adiabatic electron. The semi-local
dispersion relation for ITG is\cite{Kim1994}{\scriptsize
\begin{equation}\label{eq:itg_kine}
    1+\frac{1}{\tau}-\Big\{\frac{2}{\sqrt{\pi}}\int_{0}^{\infty}dv_\perp \int_{-\infty}^{\infty}dv_\parallel\frac{[\omega-\omega_{T}(v)]J_0^2(\sqrt{2}k_\perp v_\perp)}
    {[\omega-k_\parallel v_\parallel-\omega_{D}(v)]}v_\perp e^{-v^2}\Big\}=0,
\end{equation}}
where we have used the normalization $k_\perp\to k_\perp\rho_i$,
$k_\parallel\to \sqrt{2}k_\parallel L_n$, $\omega\to
\omega/(v_{ti}/L_n)$, $v\to v/(\sqrt{2}v_{ti})$. Thus, after
normalization, $b_i=k_\perp^2$,
$\omega_{T}(v)=\omega_{*i}\Big[1+(v_\perp^2+v_\parallel^2-\frac{3}{2})\eta_i\Big]$,
$\omega_D(v)=2\omega_{di}(v_\parallel^2+
    v_\perp^2/2)$, and
$\omega_{*i}=k_\perp$, $\omega_{di}=\omega_{*i}L_n/R=\epsilon_n
k_\perp$. In ballooning space, $k_\perp^2=k_\theta^2(1+s^2\eta^2)$,
$\partial/\partial\eta\to iqRk_\parallel$, and
$\omega_D(v,\eta)=2\epsilon_n\omega_{*i}(\cos\eta+s\eta\sin\eta)(v_\perp^2/2+v_\parallel^2)$.

\begin{table}[!h]\label{tab:semilocal_par}
\begin{center}\caption{Parameters for semi-local model.}
\begin{tabular}{c|ccc}
\hline\hline - & $\langle k_\parallel^2\rangle/k_c^2$  &  $\langle
k_\perp^2\rangle/k_\theta^2$  & $\langle
\omega_D\rangle/(2\epsilon_n\omega_{*i})$
\\\hline
$l=0$  &  1/3  & $1+3s^2/4$   & (1+0.75s)/exp(3/8)  \\
$l=1$ & 1   &  $1+9s^2/4$  &  (4+27s)/16/exp(3/8)\\
~~$l=2$ ~~ & ~~~81/55~~ &  ~~$1+5s^2/12$~~  & ~~(228+91s)/192/exp(3/8) \\
\hline\hline
\end{tabular}
\end{center}
\end{table}

For semi-local theory, we use trial function of mode structure
$\phi(\eta)$ [i.e., $f(\eta)$ in Sec.\ref{sec:model}] and calculate
the local parameters by\cite{Hirose1995}
\begin{equation}\label{eq:kpar_avg}
    \langle k_\parallel^2\rangle=-k_c^2{\int\phi^*\frac{d^2\phi}{d\eta^2} d\eta }/{\int\phi^*\phi d\eta} ,
\end{equation}
\begin{equation}\label{eq:kperp_avg}
    \langle k_\perp^2\rangle=k_\theta^2{\int\phi^*(1+s^2\eta^2)\phi d\eta }/{\int\phi^*\phi d\eta} ,
\end{equation}
\begin{equation}\label{eq:wd_avg}
    \langle \omega_D\rangle={\int\phi^*[2\epsilon_n\omega_{*i}(\cos\eta+s\eta\sin\eta)]\phi d\eta }/{\int\phi^*\phi d\eta} ,
\end{equation}
where $k_c=1/qR$ and we will use $k_\parallel=\sqrt{\langle
k_\parallel^2\rangle}$ to do the calculation. The mode structures
for different eigenstates $l$ are chosen based on the analytical
solution in Sec.\ref{sec:anly} as Hermite functions
$f_l(\eta)=H_l(\sqrt{-h}\eta)e^{-\sqrt{-h}\eta^2}$, where we set
$\sqrt{-h}=1/3$. The corresponding $\langle k_\parallel^2\rangle$,
$\langle k_\perp^2\rangle$ and $\langle \omega_D\rangle$ are
calculated in Table II for $l=0,1,2$. For these parameters, the
above semi-local model can give similar $l=0$ ITG solutions v.s.
$k_\theta\rho_i$ as those in the Fig.1 in Ref.\cite{Rewoldt2007}.
Actually, the above approach [e.g., solving $\phi(\eta)$ from
theoretical model and use it to calculate $\langle
k_\parallel^2\rangle$] is also used to study the seed parallel
Reynolds stress\cite{Xie2012} and parallel momentum
transport\cite{Lu2015}.

Fig.\ref{fig:kinetic_jump} shows a typical numerical solutions of
Eq.(\ref{eq:itg_kine}). The velocity integral is numerical
calculated by adaptive Simpson approach. The semi-quantitative
critical gradient jump of the most unstable mode from $l=0$ to $l=1$
is $\epsilon_n<\epsilon^c\sim 0.06$ and thus the critical
temperature gradient parameter $RL_T^{-1}=\eta_i/\epsilon^c\sim50$
(note that we have fixed $\eta_i$, thus we have not
distinguished the separate effects of the density  gradient or
temperature gradient). This value is close to the simulation jump
gradient in Ref.\cite{Xie2015a}, i.e., $RL_T^{-1}=40-80$. This
gradient is quite large and mainly exist at edge regions.


The above simplified calculations also show very similar jump
behavior as that by a more comprehensive 1D scanning\cite{Han2016}
of parameters for unconventional ITGs using HD7\cite{Dong2004} code.
Considering that to obtain the quantitative (especially the global)
critical gradient is still challenging due to
sensitive of numerical model as mentioned in Sec.\ref{sec:intro},
those discussions are out of the scope of the present study.

\section{Summary and discussion}\label{sec:summ}

In this work, we solve a global 2D drift wave model equation to
understand the general features of drift wave in steep gradient,
which is particular important to understand the edge plasmas physics
(e.g, the high confinement mode). Analytical solutions in fluid
limit for 1D ballooning and 2D Fourier are compared and agree
closely with numerical solutions. The unconventional drift modes can
be understood by two parameters: $\vartheta_k$ (leads to
quasi-continuous change of frequency and envelop or phase of mode
structure) and $l_\eta$ (leads to discontinuous jump of frequency and
high order harmonic of $\delta\phi_m$). The steep gradient profile
can largely change the local solution by causing both deviation of
frequency and twisting of mode structure. The theory can also
explain the twisting direction, which is determined by the imaginary
part of $b=k_\theta^2s^2\omega_{xx}\omega_{\vartheta_k\vartheta_k}$.
To give a more accurate calculation of the critical jump gradient,
we also show the kinetic solutions of a semi-local model with
velocity space integral, which gives critical jump temperature
parameter $RL_T^{-1}\sim50$ and are close to the gyrokinetic
simulation value $RL_T^{-1}=40-80$ in Ref.\cite{Xie2015a}.

The present work may be considered as a starting point to understand
the drift wave in steep gradient. And future works can include:
using more accurate kinetic model, calculating quasi-linear
diffusion, studying the linear and nonlinear consequences. However,
those works may not be straightforward. For example, the more
accurate kinetic models are usually difficult to obtain complete
solutions due to the lacking of powerful numerical approach or
limitation of computation time. Two possible approaches have been used
to solve kinetic model with complete solutions in given complex
domain: Nyquist contour integral method (cf. GLOGYSTO solver in
\cite{Brunner1998}) and transformation method base on Pad\'e
approximation (cf. PDRK solver in \cite{Xie2016}). The quasi-linear
study in Ref.\cite{Horton1981} should also be extended by including
global mode structure variation. Although the model used here is
simple and analytical approach here is standard, the physical
understanding behind is not that trivial as first glance. We should
also emphasize that the linear physics here is merely a first step
to understand the future study of the more important and interesting
nonlinear consequences (preliminary studies can be found at, cf.
Ref.\cite{Xie2015b,Pueschel2015}).

\acknowledgments Discussions and communications with D. R. Ernst, H.
T. Chen, C. J. McDevitt, D. Dickinson, T. Xie, Z. X. Lu, L. Chen, Y.
Xiao and J. Q. Li are acknowledged. The work was supported by the
China Postdoctoral Science Foundation No. 2016M590008 and the
ITER-China Grant No. 2013GB112006.

\appendix
\section{Polynomial form}\label{sec:poly_weber}

The polynomial form can tell us how many solutions exist in the
system and also all of them can be obtained by standard numerical
approach.

\subsection{1D analytical solution}
Polynomial form of Eq.(\ref{eq:itg1d_limit_3}) is
\begin{equation}\label{eq:itg1d_poly}
\sum_j a_j \omega^j=0,
\end{equation}
where, for $\vartheta_k=0$, $j=0,1,\cdots,5$, $a_0=(1+2 l)^2 \alpha
_2 \epsilon_n^{-2} \eta _s^2$, $a_1=\epsilon_n^{-1} \eta _s \{2 (1+2
l)^2 \alpha _2+k^2 [s^2(1+2 l )^2+q^2 \alpha _3^2] \epsilon_n^{-1}
\eta _s\}$, $a_2=(1+2 l)^2 \alpha _2+2 k^2 \epsilon_n^{-1} \eta _s
[s^2(1+2 l )^2+q^2 \alpha _3 (-\epsilon_n^{-1} +\alpha _3+k^2
\epsilon_n^{-1} \eta _s)]$, $a_3= k^2s^2(1+2 l )^2+k^2q^2
\epsilon_n^{-2}+k^2q^2 \{\alpha _3^2+k^2 \epsilon_n^{-2} \eta _s (-2
 +k^2 \eta _s)+2 \alpha _3
\epsilon_n^{-1}[-1 +(1+2 k^2-i\delta_e) \eta _s]\}$, $a_4=2 k^2 q^2
(1+k^2-i\delta_e) (-\epsilon_n^{-1} +\alpha _3+k^2 \epsilon_n^{-1}
\eta _s)$, $a_5=k^2 q^2 (1+k^2-i\delta_e){}^2$. For
$\vartheta_k\neq0$, $j=0,1,\cdots,7$, the coefficients $a_j$ can
also be obtained straightforwardly, but are too long. Thus we do not
list them here.

\subsection{2D equation}

Polynomial form of Eq.(\ref{eq:itg2d}) is
\begin{eqnarray}\label{eq:itg2dx_p3}\nonumber
    &&\omega^3\Big[k^2s^2\frac{d^2}{dz^2}+i\delta_e-
    k^2-1\Big]\phi_m(z)+\\
    &&\omega^2\epsilon_n^{-1}\Big[\eta_sk^2s^2\frac{d^2}{dz^2}-\eta_sk^2+1\Big]\phi_m(z)+\\\nonumber
    &&\omega\Big[\frac{1}{k^2q^2}(z-m)^2\Big]\phi_m(z)+\epsilon_n^{-1}\Big[\eta_s\frac{1}{k^2q^2}(z-m)^2\Big]\phi_m(z)-\\\nonumber
    &&\omega^2\chi\Big[(1+s\frac{d}{dz})\phi_{m+1}(z)+(1-s\frac{d}{dz})\phi_{m-1}(z)\Big]-\\
    &&\omega\eta_s\chi\epsilon_n^{-1}\Big[(1+s\frac{d}{dz})\phi_{m+1}(z)+(1-s\frac{d}{dz})\phi_{m-1}(z)\Big]=0,\nonumber
\end{eqnarray}
where $\epsilon_n^{-1}=\epsilon_n^{-1}(x)$. The above equation is
solved numerically in the article with zero boundary condition.


\begin{thebibliography}{99}

\bibitem{Horton1999} W. Horton, Rev. Mod. Phys., \textbf{71}, 735 (1999).

\bibitem{Pearlstein1969} L. D. Pearlstein and H. L. Berk, Phys. Rev. Lett.,
\textbf{23}, 220 (1969).

\bibitem{Chen1980} L. Chen and C. Z. Cheng, Phys. Fluids, \textbf{23}, 2242 (1980).

\bibitem{Horton1981} W. Horton, D. Choi and W. M. Tang, Phys. Fluids, \textbf{24},
1077 (1981).

\bibitem{Xie2015a} H. S. Xie and Y. Xiao, Phys. Plasmas, \textbf{22}, 090703 (2015).


\bibitem{Ernst2005} D.R. Ernst, K. Zeller, N. Basse, L. Lin, M. Porkolab, W. Dorland,
and A. Long, Bull. Am. Phys. Soc., \textbf{50} 235 (2005).
http://www-internal.psfc.mit.edu/research/alcator/
pubs/APS/APS2005/ernst.pdf

\bibitem{Wang2012} E. Wang, X. Xu, J. Candy, R. Groebner, P. Snyder, Y.
Chen, S. Parker, W. Wan, G. Lu, and J. Dong, Nucl. Fusion, 52,
103015 (2012).

\bibitem{Rewoldt2007}  G. Rewoldt, Z. Lin and Y. Idomura, Computer Physics Communications,
\textbf{177}, 775 (2007).

\bibitem{Candy2005}  J. Candy, Phys. Plasmas, \textbf{12}, 072307
(2005).

\bibitem{Fulton2014} D. P. Fulton, Z. Lin, I. Holod and Y. Xiao, Phys. Plasmas, \textbf{21}, 042110 (2014).

\bibitem{Liao2016}
X. Liao, Z. Lin, I. Holod, Y. Xiao, B. Li, and P. B. Snyder,
``Microturbulence in DIII-D tokamak pedestal. III. Effects of
collisions", Phys. Plasmas, in press, (2016).

\bibitem{Xie2012} T. Xie, Y. Z. Zhang, S. M. Mahajan and A. K. Wang,
Phys. Plasmas, \textbf{19}, 072105 (2012).

\bibitem{Dickinson2014} D. Dickinson, C. M. Roach, J. M. Skipp and H. R.
Wilson, Phys. Plasmas, \textbf{21}, 010702 (2014).

\bibitem{Connor1987} J. W. Connor and J. B. Taylor, Phys. Fluids \textbf{30}, 3180
(1987).


\bibitem{Singh2014} R. Singh, S. Brunner, R. Ganesh and F. Jenko, Phys. Plasmas, \textbf{21}, 032115
(2014).

\bibitem{Lu2015} Z. X. Lu, Phys. Plasmas, \textbf{22}, 052118 (2015).

\bibitem{Abdoul2015} P. A. Abdoul, D. Dickinson, C. M. Roach and H. R. Wilson, Plasma Phys. Control. Fusion, 57,
065004 (2015).

\bibitem{Hastie1979} R. Hastie, K. Hesketh and J. Taylor, Nucl.
Fusion, 19, 1223 (1979).

\bibitem{Taylor1993} J. B. Taylor, J. Connor and H. R. Wilson, Plasma Phys. Control Fusion, \textbf{35},
1063 (1993).

\bibitem{Deng2010} W. Deng, Z. Lin, I. Holod, X. Wang, Y. Xiao and W. Zhang,
Phys. Plasmas, \textbf{17}, 112504 (2010).

\bibitem{Wang2010} X. Wang, F. Zonca, and L. Chen, Plasma Phys. Controlled Fusion \textbf{52},
115005 (2010).

\bibitem{Zhang2010} H. Zhang, Z. Lin, I. Holod, X. Wang, Y. Xiao, and W. Zhang, Phys.
Plasmas \textbf{17}, 112505 (2010).

\bibitem{Bass2013} E. M. Bass and R. E. Waltz, Phys. Plasmas \textbf{20}, 012508 (2013).

\bibitem{Ma2015} R. Ma, F. Zonca and L. Chen, Physics of Plasmas,
\textbf{22}, 092501 (2015).

\bibitem{Bottino2004} A. Bottino, A. G. Peeters, O. Sauter, J. Vaclavik, L. Villard, and
ASDEX Upgrade Team, Phys. Plasmas \textbf{11}, 198 (2004).

\bibitem{Connor1978} J. W. Connor, R. J. Hastie and J. B. Taylor, Phys. Rev. Lett.,
, \textbf{40}, 396 (1978).

\bibitem{Connor1979} J. W. Connor, R. J. Hastie and J. B. Taylor, Proc. R. Soc. London
Ser.A, \textbf{365}, 1 (1979).

\bibitem{Romanelli1993} F. Romanelli and F. Zonca, Phys. Fluids B \textbf{5}, 4081 (1993).

\bibitem{Dong1997} J. Q. Dong, S. M. Mahajan and W. Horton, Phys. Plasmas, \textbf{4},
755 (1997).

\bibitem{Ernst2004} D. R. Ernst, P. T. Bonoli, P. J. Catto, W. Dorland, C. L. Fiore, R. S. Granetz, M. Greenwald, A. E. Hubbard, M.
Porkolab, M. H. Redi, J. E. Rice, K. Zhurovich and Alcator C-Mod
Group, Phys. Plasmas, \textbf{11}, 2637 (2004).

\bibitem{Zhang1991} Y. Z. Zhang and S. M. Mahajan, Phys. Lett. A \textbf{157}, 133 (1991).

\bibitem{Xie2016b} T. Xie, H. Qin, Y. Z. Zhang and S. M. Mahajan,  Phys. Plasmas, \textbf{23}, 042514
(2016).

\bibitem{Heading1962} J. Heading, Q. J. Mech. Appl. Math. \textbf{15}, 215 (1962).

\bibitem{Migliano2013} P. Migliano, Y. Camenen, F. J. Casson, W. A. Hornsby, and A. G.
Peeters, Phys. Plasmas \textbf{20}, 022101 (2013).

\bibitem{Kim1994} J. Y. Kim, Y. Kishimoto, W. Horton and T. Tajima, Phys. Plasmas, \textbf{1},
927 (1994).

\bibitem{Hirose1995} A. Hirose, L. Zhang and M. Elia, Phys. Plasmas, \textbf{2},
859 (1995).

\bibitem{Han2016} M. K. Han, Z. X. Wang, J. Q. Dong and H. R. Du, Ion
temperature gradient modes of unconventional ballooning structures
in pedestal region of tokamaks, to be submitted (2016).


\bibitem{Dong2004} J. Q. Dong, L. Chen, F. Zonca and G. D. Jian, Phys. Plasmas, \textbf{11}, 997 (2004).

\bibitem{Brunner1998} S. Brunner, M. Fivaz, T. M. Tran and J. Vaclavik, Phys. Plasmas, \textbf{5},
3929 (1998).

\bibitem{Xie2016} H. S. Xie and Y. Xiao, Plasma Science and Technology,
\textbf{18}, 97 (2016).

\bibitem{Xie2015b} H. S. Xie, Numerical Simulations of Micro-turbulence in Tokamak
Edge, PhD thesis, Zhejiang University, (2015).
http://hsxie.me/files/thesis/

\bibitem{Pueschel2015} M. J. Pueschel,  D. R. Hatch, P. W. Terry and J.W.
Connor, Edge Turbulence: Mode Parity and Consequences for Transport,
U.S. Transport Task Force Workshop, Salem, May, 2015.

\end{thebibliography}
\end{document}